# Multiple stochastic resonances and inverse stochastic resonances in asymmetric bistable system under the ultra-high frequency excitation

Cong Wang[1], Zhongqiu Wang[2], Jianhua Yang[1,*], Miguel A. F. Sanjuán[3], Gong Tao[1], Zhen Shan[1], Mengen Shen[1]

[1] *Jiangsu Key Laboratory of Mine Mechanical and Electrical Equipment, School of Mechatronic Engineering, China University of Mining and Technology, Xuzhou 221116, P. R. China*

[2] *School of Computer Science and Technology, China University of Mining and Technology, Xuzhou 221116, P. R. China*

[3] *Nonlinear Dynamics, Chaos and Complex Systems Group, Departamento de Física, Universidad Rey Juan Carlos, Tulipán s/n, 28933 Móstoles, Madrid, Spain*

**Abstract** Ultra-high frequency linear frequency modulation (UHF-LFM) signal, as a kind of typical non-stationary signal, has been widely used in microwave radar and other fields, with advantages such as long transmission distance, strong anti-interference ability, and wide bandwidth. Utilizing optimal dynamics response has unique advantages in weak feature identification under strong background noise. We propose a new stochastic resonance method in an asymmetric bistable system with the time-varying parameter to handle this special non-stationary signal. Interestingly, the nonlinear response exhibits multiple stochastic resonances (MSR) and inverse stochastic resonances (ISR) under UHF-LFM signal excitation, and some resonance regions may deviate or collapse due to the influence of system asymmetry. In addition, we analyze the responses of each resonance region and the mechanism and evolution law of each resonance region in detail. Finally, we significantly expand the resonance region within the parameter range by optimizing the time scale, which verifies the effectiveness of the proposed time-varying scale method. The mechanism and evolution law of MSR and ISR will provide references for researchers in related fields.

---

* e-mail address: jianhuayang@cumt.edu.cn (corresponding author)

# 1 Introduction

Ultra-high frequency linear frequency modulation (UHF-LFM) signal, as a kind of special non-stationary signal [1], has advantages of long transmission distance, low transmission power, strong anti-interference ability and wide bandwidth. This signal has been widely used in wireless communication [2] and high-precision measurement technology [3, 4], etc. In the system, the radar can modulate the chirp signal to obtain a larger time-wide bandwidth, where higher range resolution and larger range are realized. The analysis of UHF-LFM signal focuses on the detection and estimation of the frequency modulation slope and initial frequency, while the time-frequency analysis is widely used [5, 6]. These methods mainly include wavelet transform [7], Wigner-Ville distribution [8], fractional Fourier transform [9, 10], etc. However, the signal frequency estimation ability of the time-frequency analysis method is limited due to the inevitable presence of noise. The filtering method detects the signal aiming at noise elimination, but the noise reduction may remove the useful characteristic signal and affect the detection ability of the radar [11, 12]. Different from the above methods, stochastic resonance (SR) [13] method can utilize noise energy to amplify weak signals with the synergistic effect of system, signal and noise, where noise is individually removed.

With further exploration, the theory of SR has been widely studied and applied in the fields of physics [14], chemistry [15], mechanics [16], medicine [17, 18], neural networks [19], and various other fields. Furthermore, many scholars have further studied the resonance response under different excitations, and expanded different system models, such as monostable systems [20], bistable systems [21], metastable systems [22, 23], and asymmetric systems [24-27]. For instance, Shi et al. [25] discussed SR in a time polo-delayed asymmetry bistable system driven by multiplicative white noise and additive color noise. Different from the common single resonance response, sometimes the system also exhibits complex characteristics with the existence of multiple stochastic resonances (MSR) [28-32]. Yu et al. [28] found that the time delay can induce MSR due to the transition of network dynamics induced by time delays. Besides, the phenomenon of inverse stochastic resonances (ISR) [33-35] has also been specifically analyzed. Li et al. [34] mainly studied the phenomenon of ISR induced by non-Gaussian noise in a representative Hodgkin–Huxley model. Mi et al. [35] quantified the occurrence of SR by obtaining response time series and found that the Lévy noise intensity can cause both SR and ISR. The enrichment and essential nature of investigating the resonance phenomenon under various signals and system excitations are evident.

There have been many explorations on how to enhance signal features via SR [36-38]. However, the LFM signal has a wide bandwidth, which brings difficulties to signal processing [39], especially for UHF-LFM signal [40]. The traditional re-scaled method with fixed-scale coefficient cannot adapt to the time-varying frequency of LFM signal, and cannot obtain the ideal resonance effect. To achieve high-quality processing of non-stationary signals by SR, some re-scaled methods have been proposed, such as the sliding window method [41], the frequency-shifted method [42,43], the normalized re-scaled method [44], and the general re-scaled method [45]. The re-scaled method with fixed scale is relatively rough and cannot achieve the strongest resonance required for all frequency bands. For UHF-LFM signals, it is necessary to propose a re-scaled method that is able to be better adapt frequency transformation.

Despite extensive investigations, there remains a deficiency in research concerning the resonance of an asymmetric bistable system under UHF-LFM signal excitation. It is significant for studying the strongest dynamic response of nonlinear systems and clarifying the response characteristics of the resonance region. Inspired by these issues, we propose a time-varying re-scaled method to obtain the strongest resonance effect. Furthermore, we provide a specific explanation of the underlying mechanism and evolutionary law of dynamic response under UHF-LFM signal excitation.

Our research mainly contains the following parts: in section 2, we give an asymmetric system with time-varying scale for processing UHF-LFM signal. In section 3, the nonlinear response patterns of the asymmetric bistable system under UHF-LFM signal excitation are studied, and the inducing parameters and internal mechanisms of MSR and ISR are analyzed. The asymmetric parameter and scale bases of the system are optimized based on the resonance regions. The main findings of this paper are discussed at the end in the conclusions section.

## 2 The model

Here, we first introduce the specific composition of the nonlinear model and typical characteristics of the UHF-LFM signal in asymmetric bistable system. In addition, the system equation is rebuilt through the time-varying re-scaled method.

## 2.1 Time-varying dynamic model

The time-varying asymmetric bistable system that is excited by UHF-LFM signal is

$$\frac{dx(t)}{dt} = -\frac{\partial U(x)}{\partial x} + s(t) + N(t). \tag{1}$$

Herein, $U(x)$ denotes the potential function of the time-varying system. The input signal $s(t)$ is the UHF-LFM signal, and $N(t)$ is a Gaussian white noise.

In equation (1), an asymmetric potential function is expressed as

$$U(x) = -\frac{1}{2}a(t)x^2 + \frac{1}{4}b(t)x^4 + \frac{1}{3}c(t)x^3. \tag{2}$$

The potential of the asymmetric bistable system is shown in Fig. 1.

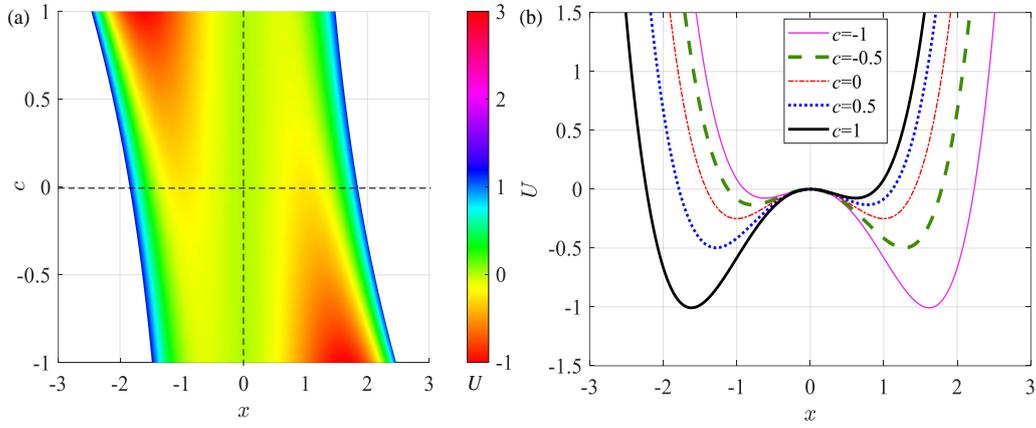

**Fig. 1.** The role of the asymmetric parameter $c$ on the potential configuration. (a) The potential curve varies with the asymmetric parameter $c$ in the two-dimensional plane. (b) The potential curves under 5 cases of the asymmetric parameter $c$.

In Fig. 1(a), we present the variation in the potential function versus the asymmetric parameter $c$ in a two-dimensional plane. To accentuate the asymmetry introduced by the asymmetric parameter, we chose five sets of potential function curves with different $c$ for further explanation. When asymmetric parameter $c<0$, as its absolute value increases, the potential well of the right potential well increases and the potential well becomes deeper, while the barrier of the left potential well decreases and the potential well becomes shallower. When asymmetric parameter $c>0$, with the increase of $c$, the potential well of the left potential well increases and the potential well becomes deeper, while the potential well of the right potential well decreases and the potential well becomes shallower. In particular, when asymmetric parameter $c=0$, it is a potential function of a symmetric bistable system.

The frequency of the UHF-LFM signal $s(t)$ increases linearly with time and is

expressed as

$$f = f_c + \frac{B}{T}t, \tag{3}$$

the signal frequency starts with $f_c$. The parameter $T$ is the time width of the chirp signal, and $B$ is the bandwidth of the chirp signal.

The equation for the UHF-LFM signal is

$$s(t) = A\sin\left(2\pi f_c + \pi \frac{B}{T}t^2\right). \tag{4}$$

Referring to the mainstream millimeter-wave radar [46], the parameters in equation (4) are selected as $A$=0.5, $f_c$=77GHz, $T$=0.03 microsecond.

To simulate the UHF-LFM signal that is submerged by noise, we build a noisy signal $s(t)$ with *SNR* of -27dB. The time domain and time-frequency domain of the noisy signal $s(t)$ are presented in Fig. 2.

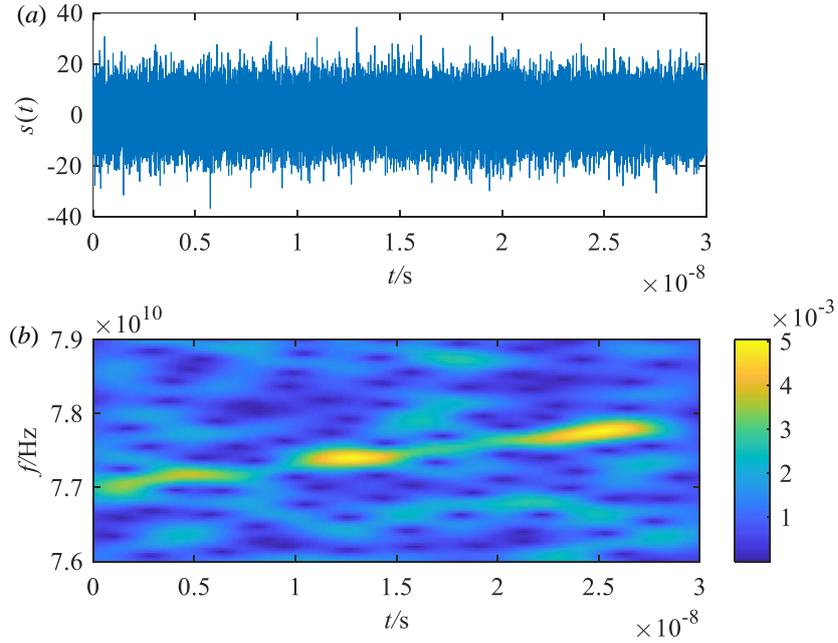

**Fig. 2.** Noisy signal $s(t)$. (a) Time domain of noisy signal $s(t)$, (b) time-frequency domain of noisy signal $s(t)$.

## 2.2 Theoretical formulation

We propose the time-varying re-scaled method to make the parameters $a(t)$, $b(t)$, $c(t)$ and the input $s(t)$ meet the optimal parameter match to achieve SR. In the time-varying re-scaled method, the scale with time-varying coefficient is defined as $\lambda(t)$

$$\lambda(t) = \lambda_0 f(t), \tag{5}$$

where $\lambda_0$ is the base of the time-varying scale, and $f(t)$ is the frequency of the UHF-LFM signal.

Then, the following parameter matching relationships are obtained

$$a(t) = \lambda(t)a_1, \quad b(t) = \lambda(t)b_1, \quad c(t) = \lambda(t)c_1. \tag{6}$$

Among them, $a_1$, $b_1$, and $c_1$ are small parameters. Introducing the variable substitution $x=z(v)$, $v=\lambda(t)t$, substituting equation (5) into equation (1), one gets

$$\frac{dz(v)}{dv} = \frac{a(t)}{\lambda(t)}z(v) - \frac{b(t)}{\lambda(t)}z^3(v) - \frac{c(t)}{\lambda(t)}x^2(v) + \frac{1}{\lambda(t)}s(\frac{v}{\lambda(t)}) + \frac{1}{\lambda(t)}N(\frac{v}{\lambda(t)}). \tag{7}$$

Substituting equation (7) into equation (8), one obtains

$$\frac{dz(v)}{dv} = a_1 z(v) - b_1 z^3(v) - c_1 x^2(v) + \frac{1}{\lambda(t)}s(\frac{v}{\lambda(t)}) + \frac{1}{\lambda(t)}N(\frac{v}{\lambda(t)}). \tag{8}$$

Obviously, equation (8) satisfies the small parameter condition of classic SR.

Comparing equation (8) with equation (1), it is observed that the amplitude of the input signal $s(t)$ and the noise strength of $N(t)$ in equation (8) are scaled to $1/\lambda(t)$ times of that in equation (1). By restoring the amplitude of the input $s(t)$ and the noise strength of $N(t)$ in equation (8) to the original signal amplitude, it is obtained that

$$\frac{dz(v)}{dv} = a_1 z(v) - b_1 z^3(v) - c_1 x^2(v) + s(\frac{v}{\lambda(t)}) + N(\frac{v}{\lambda(t)}). \tag{9}$$

When equation (9) satisfies the condition of SR, equation (10) must also satisfy the condition of SR,

$$\frac{dx(t)}{dt} = a(t)x(t) - b(t)x^3(t) - c(t)x^2(t) + \lambda(t)s(t) + \lambda(t)N(t). \tag{10}$$

Among them, $a(t)$, $b(t)$, and $c(t)$ are all time-varying system parameters, and equation (10) can achieve SR for the UHF-LFM signal.

## 3 Resonance response analysis of the system

We investigate here the influence of asymmetric parameter on the nonlinear response of the asymmetric bistable system, quantifying the response characteristics of the resonance region using cross-correlation coefficient as an indicator. By analyzing and explaining the phenomena of MSR and ISR in detail, combined with the response characteristics of asymmetric parameter and the optimal time-varying scale cardinality $\lambda$ analysis. Finally, the enhancement of noisy UHF-LFM signals is achieved.

## 3.1 The influence of asymmetric parameter on the response characteristics of MSR and ISR

Due to the property of the time-varying frequency, the frequency of the LFM signal is not concentrated on one or more countable frequencies. Therefore, the commonly used indices such as signal-to-noise ratio, signal-to-noise ratio gain, and spectral amplification parameter are difficult to apply to measure non-stationary signals. Fortunately, the cross-correlation coefficient which characterizes the similarity of waveforms between the input and the output in the time domain is an effective index [47], which is defined as

$$C_s = \frac{\sum_{t=1}^{n}(s(t)-\bar{s})(x(t)-\bar{x})}{\sqrt{\sum_{t=1}^{n}(s(t)-\bar{s})^2 \sum_{t=1}^{n}(x(t)-\bar{x})^2}}. \tag{12}$$

For the signal presented in Fig. 1, a pseudo three-dimensional plot of the cross-correlation coefficient versus the system parameters $a_1$ and $b_1$ under different asymmetric parameters are given in Fig.3. The larger value of the cross-correlation coefficient, the higher similar of the response to the input signal, and leading to the strong SR. It is found that under some specific conditions, the average firing rate exhibits a minimum as the noise level varies. Such new phenomenon of inhibition effect is so-called ISR. Specifically, there are multiple resonance regions in the system response of Fig. 3, including bright SR regions and dark ISR regions.

The asymmetric parameters of the potential function in this paper induce the formation of two stable equilibrium points and one unstable equilibrium point. Asymmetry causes changes in the depth of the potential well, and the resonance response is closely related to the potential well where motion occurs [47, 48]. When the asymmetric parameter remains unchanged, changes in other system parameters cause a change in the depth of the potential well, which in turn affects the motion of the response within the potential well. However, the asymmetric parameter has a greater impact on the potential well. Changing the asymmetric parameter disrupts the balance node of the system, causing the response to switch between bistable system, monostable system, and ISR system. This also causes the original resonance region to split into two parts as shown in Fig. 3 (c). They are labeled as regions ① and ②.

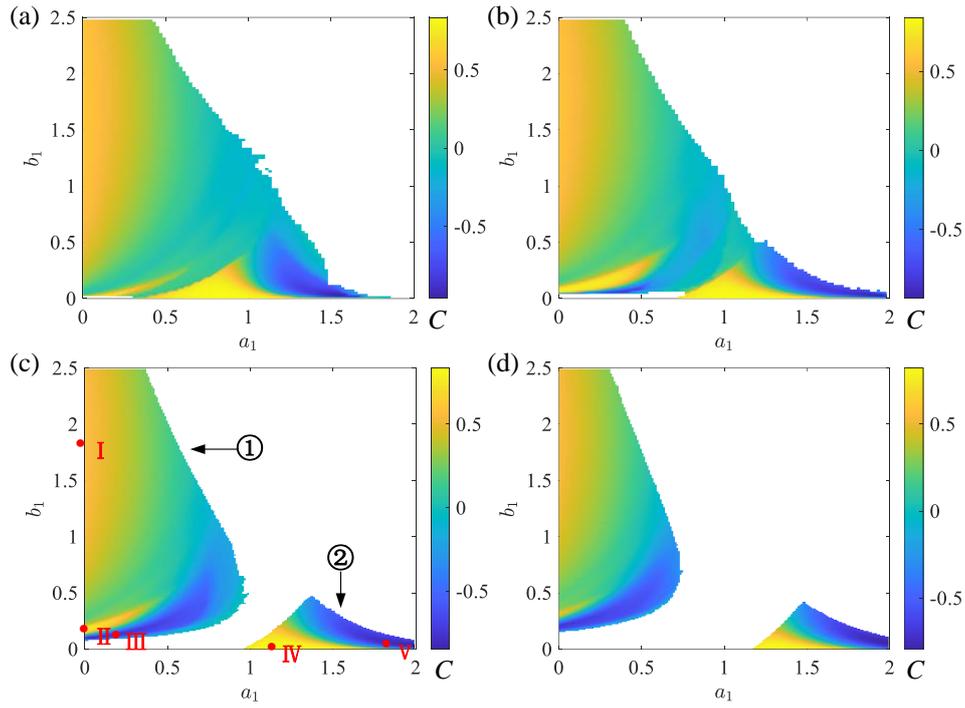

**Fig. 3.** MSR phenomenon of the asymmetric bistable system with left potential-well low and right potential-well high. The simulation parameters are $\lambda_0=20$, (a) $c_1=0.1$, (b) $c_1=0.3$, (c) $c_1=0.5$, (d) $c_1=0.7$. The color codes the value of the cross-correlation coefficient.

In Fig. 3, the system with a smaller left bias has only one resonance region, and as the asymmetry increases, the resonance region separates into two regions. In Fig. 3(a), when the asymmetric parameter is set to $c_1=0.1$, the response of the nonlinear system exhibits a SR region, where three bright SR regions have a large value and one dark ISR region has a negative cross-correlation coefficient. As the asymmetric parameter adjustment increases, the resonance region of the system tends to separate. When the asymmetric parameter is set to $c_1=0.3$, as shown in Fig. 3(b), the ISR region with negative cross-correlation coefficient appears in the lower left corner of the resonance region. At the same time, the bright SR regions and dark ISR regions on the right side of the system resonance regions gradually shift to the right. In Fig. 3(c), when the asymmetric parameter is set to $c_1=0.5$, the system resonance region splits into two resonance regions. The region of the bright SR in the lower left corner of the original resonance region gradually decreases, while the region of the dark ISR gradually expands. In Fig. 3(d), when the asymmetric parameter is set to $c_1=0.7$, the left resonance region of the system further contracts to the left, and the bright SR region in the left resonance region is basically replaced by the dark ISR region.

Meanwhile, the right resonance region further moves to the right at $a_1$=1.2.

We analyze the system response characteristics of five resonance regions in Fig. 3, reflecting the response characteristics of a time-varying asymmetric system. The response is shown in the phase trajectory in Fig. 4. Specifically, Figs. 4(a), (b), and (c) represent three SR and ISR regions in Fig. 3(c)-①. The trajectory in Fig. 4(a) shows the stable large-scale periodic state of the system response in the first bright SR region, with the state trajectory moving around the left side of the center. The system response can maintain a stable transition between two potential wells. The phase trajectory in Fig. 4(b) represents the second bright SR region. Although its cross-correlation coefficient is higher than the first bright SR region, the system only undergoes small-scale periodic state within the left potential well. The phase trajectory in Fig. 4(c) shows that the response state in the first dark ISR region exhibits the chaotic state [49], and the system output only moves in the left potential well without any transitions. Figs. 4(d) and (e) represent the bright SR region and dark ISR region on the right side of Fig. 3(c), respectively. The phase trajectory in Fig. 4(d) shows that the system response in the third bright SR region is the small-scale periodic state [50], with the output only transiting within the right potential well. The phase trajectory in Fig. 4(e) shows that the system output exhibits chaotic state in the second dark ISR region, only moving in the left potential well without any transitions, and exhibits a central symmetry relationship with the system response in Fig. 4(c).

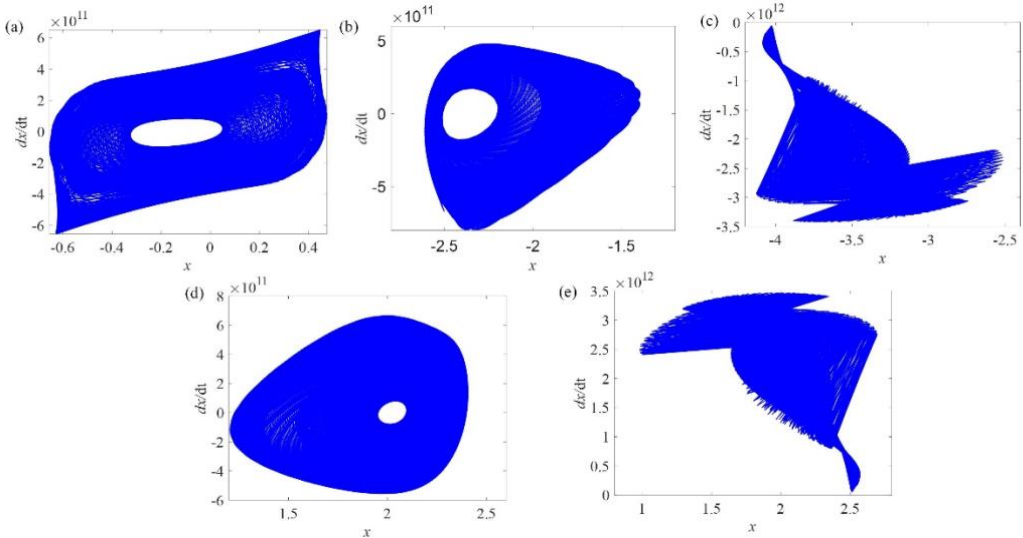

**Fig. 4.** Phase trajectory corresponding to five resonance regions in Fig. 3(c). (a) Phase trajectory of point I, (b) phase trajectory of point II, (c) phase trajectory of point III, (d) phase trajectory of point IV, (e) phase trajectory of point VII. The system parameters are $\lambda_0$=20, $c_1$=0.5.

In Fig. 4, we employ a phase diagram to analyze the motions within the five resonance regions depicted in Fig. 3(c). At point I, the depth of the potential wells on both sides of the model is uniform, enabling the motion to cross between the two potential wells with sufficient energy. This is manifested in the phase diagram as periodic oscillations between the two potential wells, corresponding to bistable system. At points II and IV, there is a significant disparity in the depth of the potential wells on either side, preventing the motion from crossing potential wells. This is depicted in the phase diagram as oscillations within a single potential well, suggesting the presence of monostable system. Notably, ISR system is exhibited at points III and V. This describes the chaotic motion under the influence of potential well depth.

When the asymmetric parameter $c<0$, the time-varying system becomes a right tilted system. As the asymmetric parameter $c$ decreases, the potential barrier of the left potential well decreases and the potential well becomes shallower, while the potential barrier of the right potential well increases and the potential well becomes deeper. Fig. 5 shows the relationship between the cross-correlation coefficient of the response versus the system parameters $a$ and $b$ when the system asymmetric parameter is negative.

When the asymmetric parameter $c<0$, the asymmetric bistable system reflects different response characteristics. As the asymmetric parameter $c$ changes, there is no separation phenomenon in the resonance region, but rather a partial collapse of the resonance region within the resonance region. By comparing the changes in the resonance region of the left tilted system in Fig. 3, it concludes that the asymmetric bistable system with left potential-well high and right potential-well low collapses in the bright SR region and the dark ISR region on the right side of the resonance region. In Fig. 5(d), when asymmetric parameter $c_1=0.7$, the right side of the system resonance region completely disappears, leaving only the three main resonance regions on the left side of the system resonance.

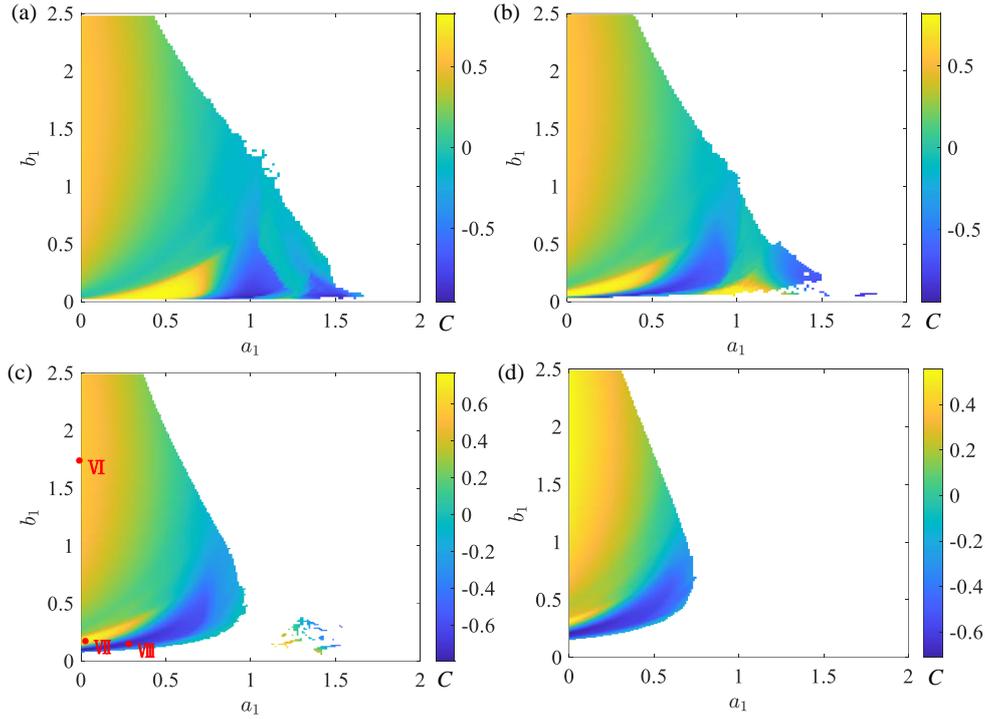

**Fig. 5.** MSR phenomenon of the asymmetric bistable system with left potential-well high and right potential-well low. The simulation parameters are $\lambda_0=20$, (a) $c_1=-0.1$, (b) $c_1=-0.3$, (c) $c_1=-0.5$, (d) $c_1=-0.7$.

In order to further study the influence of the asymmetry on the nonlinear system, we conduct a response analysis on Fig. 5(c) by the phase trajectory in Fig. 6. Compared with the left tilted system, in the first bright SR region in Fig. 6(a), the system continues to undergo a large-scale periodic state, with the output center becoming right tilted, which can also maintain transitions between potential wells. In Fig. 6(b) and (c), these two resonance regions are mainly affected by the deepening of the potential barrier on the right side of the tilted system, which is completely opposite to the movement form of the left tilted system. The resonance region is mainly affected by the right potential well, which is completely opposite to the response pattern of the system with the asymmetric parameter $c>0$.

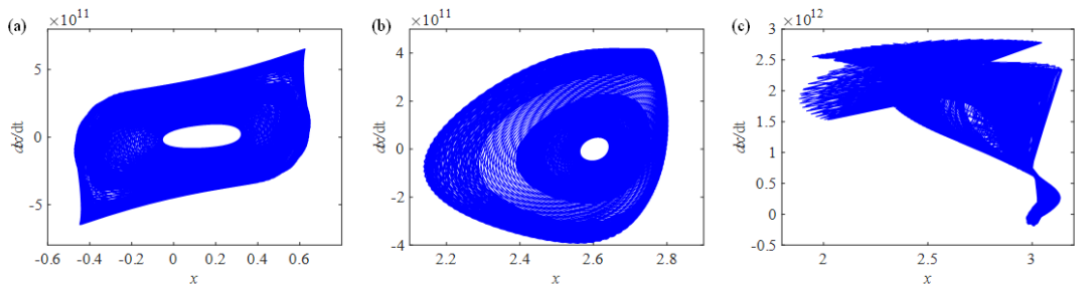

**Fig. 6.** Phase trajectory corresponding to five resonance bands in Fig. 5(c). (a) Phase

trajectory corresponding to point VI, (b) phase trajectory corresponding to point VII, (c) phase trajectory corresponding to point VIII. The system parameters are $\lambda_0=20$, $c_1=-0.5$.

By comparing Fig. 3 to Fig. 6 comprehensively, it is not difficult to find that the ISR regions and SR regions often appear in groups within the same potential well, and a whole of SR and ISR regions excited by weak potential wells in the asymmetric system move or dissipate as a whole. We define the region with large-scale periodic motion as the bistable region and the region with small-scale periodic motion as the monostable region.

For the asymmetric bistable system, the response results are not only affected by the asymmetric parameter to generate MSR and ISR, but also by the change of scale cardinality. Fig. 7 shows the relationship between system parameters $a$, $b$, and the cross-correlated coefficient $C$, as the scale cardinality $\lambda_0$ taking different values. In section 3.1, we have provided a detailed explanation of the resonance types. In Fig. 7(a), bright region only experiences periodic oscillations within a potential well. Along with the value of $c$ changing, bright region mainly occurs periodic oscillation between two potential wells in Fig. 7(c). As the scale cardinality $\lambda_0$ increases, the parameter range of the resonance region gradually narrows down. When $\lambda_0=20$, the resonance response of the system basically retains the whole bistable region.

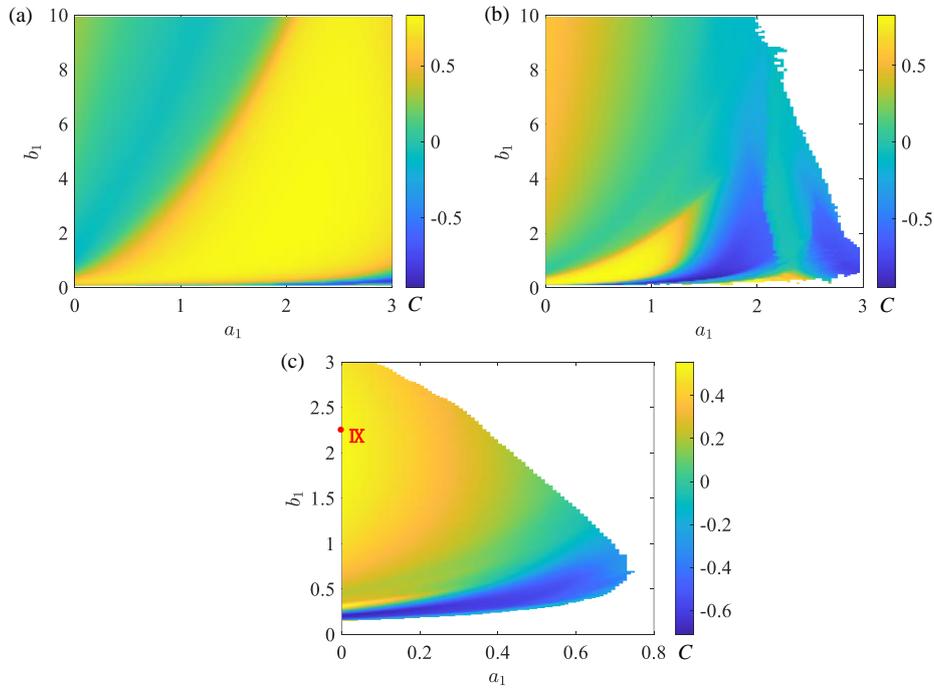

**Figure. 7.** Response at different scale cardinality $\lambda_0$. The simulation parameters are $c_1=-0.7$, (a) $\lambda_0=5$, (b) $\lambda_0=15$, (c) $\lambda_0=20$.

12 / 23

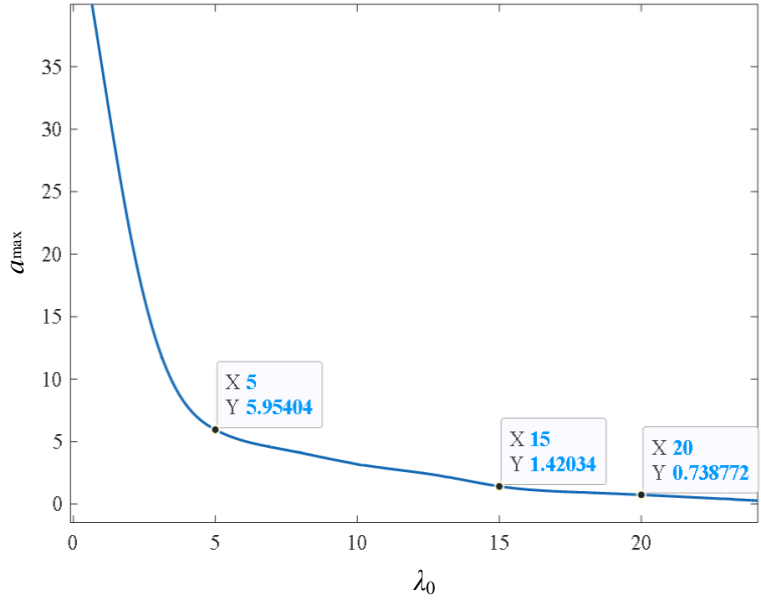

**Fig. 8.** Diagram of relationship between scale cardinality $\lambda_0$ and $a_{\max}$.

The scale cardinality $\lambda_0$ is primarily utilized to modify the frequency scale of the original signal. We can guarantee the small parameter condition of resonance phenomenon through scale transformation. As depicted in Fig. 7, the resonance region is jointly represented by parameters $a$ and $b$. We consider the maximum value of parameter $a_{\max}$ as the boundary limit for the resonance response. In Fig. 8, we plot the relationship between $a_{\max}$ and scale cardinality using experimental data. When $\lambda_0<5$, an increase in $\lambda_0$ rapidly reduces $a_{\max}$, yet it still does not satisfy the resonance criteria for small parameters. When $\lambda_0>5$, $a_{\max}$ decreases gradually with the increase in $\lambda_0$, plateauing when $\lambda_0=20$, which satisfies the necessary small parameter range for the system and effectively enhances the signal processing.

### 3.2 Time-frequency analysis of the resonance

In this section, we specifically analyzed the strongest outputs in each resonance region. Fig. 9 and Fig. 10 show the strongest output for the resonance regions in Fig. 3(c). Specifically, Figs. 9(a) and (b) respectively show the time domain and the STFT spectrum of the system strongest output corresponding to point I. The signal amplitude is significantly enhanced, while ensuring the continuity of the time-frequency characteristics. According to the strongest output of point II in Fig. 8, although the cross-correlation coefficient corresponding to point II is higher than that of point I, the response of the monostable region corresponding to point II is significantly worse than that of point I, with discontinuous waveform and severe energy divergence. In Fig. 9, although point III is located in the ISR region, the strongest output shows high consistency with the output corresponding to point II in the monostable region in both time and time-frequency domains.

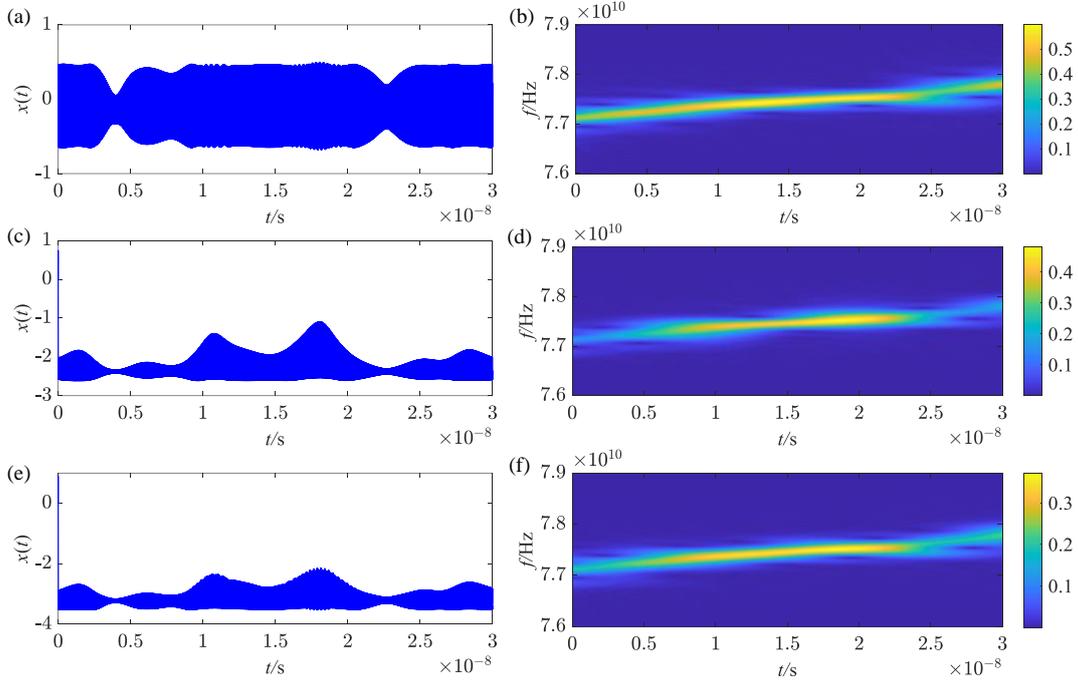

**Fig. 9.** System resonant output corresponding to the region ① in Fig. 3(c). (a) Time domain corresponding to point I, (b) the STFT spectrum corresponding to point I, (c) time domain corresponding to point II, (d) the STFT spectrum corresponding to point II, (e) time domain corresponding to point III, (f) the STFT spectrum corresponding to point III. The system parameters are $c_1$=0.5, $\lambda_0$=20.

In Fig. 10, we observe that the strongest output corresponding to points IV and V also exhibit high consistency, weak temporal continuity, and relatively divergent energy in the time-frequency domain. Based on the comprehensive analysis of Fig. 9 and Fig. 10, under the same potential well excitation, the strongest output of the system in the SR and ISR regions are consistent in both the time and frequency domains. The two sets of the SR regions and ISR regions under different potential well excitations have opposite tilted directions in the time domain, and the energy divergence of the time-frequency domain outputs is similar.

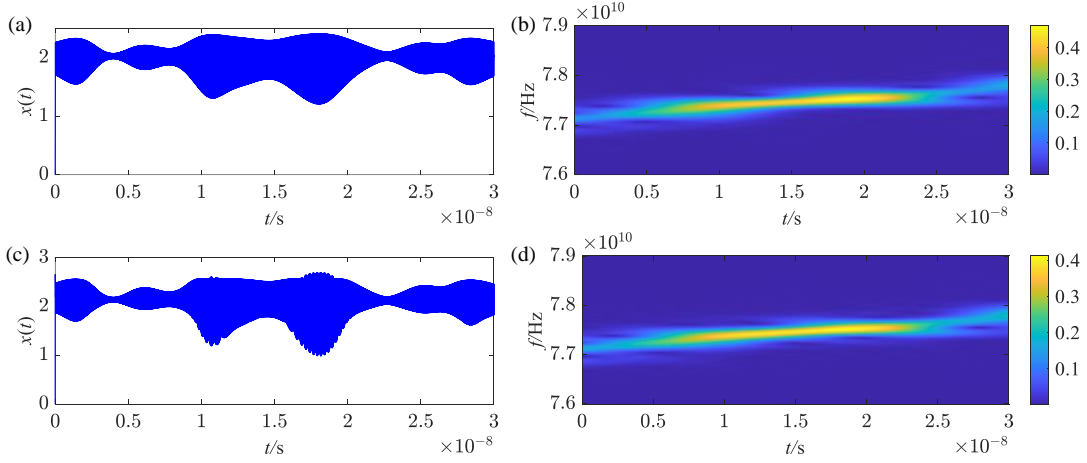

**Fig. 10.** System resonant output corresponding to the region ② in Fig. 3 (c). (a) Time domain corresponding to point IV, (b) the STFT spectrum corresponding to point IV, (c) time domain corresponding to point V, (d) the STFT spectrum corresponding to point V. The system parameters are $c_1=0.5$, $\lambda_0=20$.

To fully illustrate the output characteristics of the asymmetric system, we select the resonance regions in Fig. 5(c) for system output characteristic analysis, as shown in Fig. 11. The system still achieves the strongest output corresponding to point VI in the bistable region. Under the influence of the right potential well, the system output corresponding to points VII and VIII exhibit the same characteristics as laid in Fig. 8 and Fig. 9, and the recovery effect seems worse than the output of the bistable region.

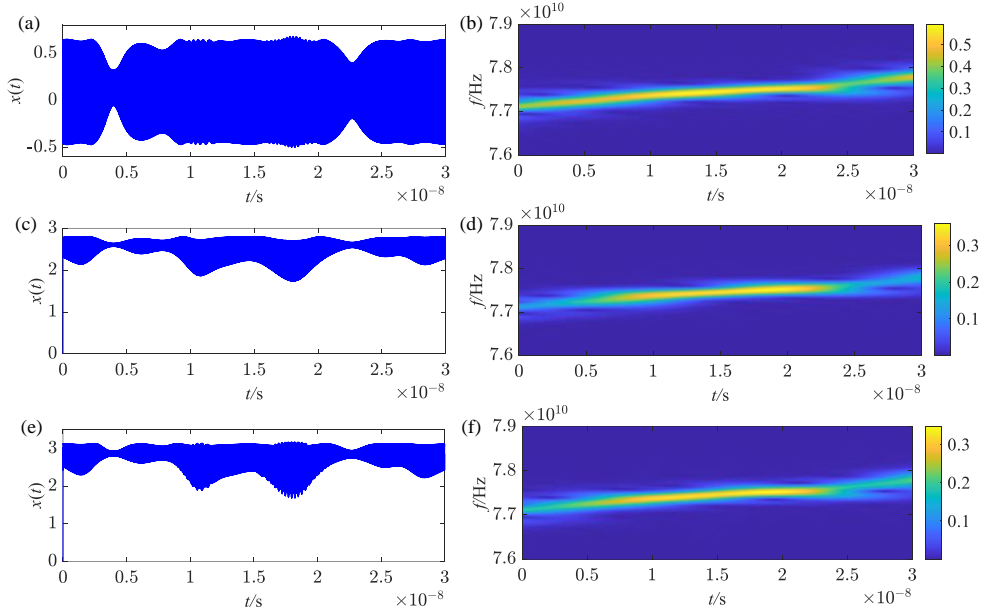

**Fig. 11.** System resonant output corresponding to the regions in Fig. 5(c). (a) Time domain corresponding to point VI, (b) the STFT spectrum corresponding to point VI, (c) time domain corresponding to point VII, (d) the STFT spectrogram corresponding to point VII, (e) time domain corresponding to point VIII, (f) the STFT spectrum

corresponding to point VIII. The system parameters are $c_1$=-0.5, $\lambda_0$=20.

When the asymmetric parameter $c$ takes a negative value, it forms a right tilted bistable model with a deep right potential-well and a shallow left potential-well. This is beneficial for actively removing non-bistable regions, reducing unnecessary analysis processes and calculation time. When $c_1$=-0.7, the resonance of the asymmetric system is mainly the bistable region, which is the region we are concerned about. Meanwhile, the increase of $\lambda_0$ gradually expands the resonance range of the system to the range of small parameters $a$ and $b$. When $\lambda_0$=20, almost all bistable regions are preserved. We bring the maximum cross-correlation coefficient point corresponding to IX of the bistable resonance regions in Fig. 7 (c) into the asymmetric system. Then, we obtained the final system resonance output and placed it in Figure 11. Comparing Fig. 2 and Fig. 12, the UHF-LFM signal is enhanced through a time-varying asymmetric bistable system, and the energy of the signal is highly concentrated in the time-frequency domain, which is clearly visible.

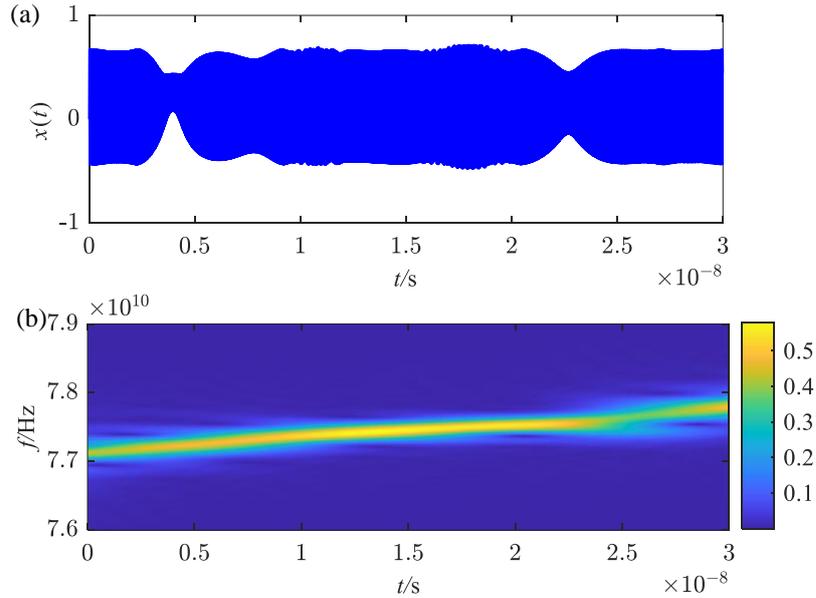

**Fig. 12.** System resonant output corresponding to the regions in Fig. 7 (c). (a) Time domain corresponding to point IX, (b) the STFT spectrogram corresponding to point IX. The system parameters are $c_1$=-0.7, $\lambda_0$=20.

The UHF-LFM non-stationary signal exhibits significance within the frequency range, rather than being concentrated at specific frequencies. Compared with the traditional re-scaled method, the time-varying re-scaled method used in this paper can better meet the characteristics of non-stationary signal. In addition, using the cross-correlation coefficient as an evaluation index, this paper analyzes the response of the asymmetric bistable system to explain the MSR and ISR under UHF-LFM signal excitation. Further, we discover the evolution and formation mechanism of MSR and ISR through novel analysis of phase trajectory.

### 3.3 Effect of noise on system response

Consider strong noise background, and we use signal-to-noise ratio($SNR$) to indicate the noise intensity in the signal. Fig. 9 compares the influences of noise on the cross-correlation coefficient for system inputs $SNR$ of -21, -23, -25, -27 and -29, respectively. With parameters $a=0.1$ and $c=0.5$, the cross-correlation coefficient $C$ exhibits consistent trends across these five $SNR$ values, indicating the adaptability of the method to varying noise intensities. Similar to Fig. 3, because the system diverges when the value of $b$ is small, an initial segment of the curve is missing data in Fig. 13. In summary, the system proposed in this paper can effectively handle signals with different noise intensities.

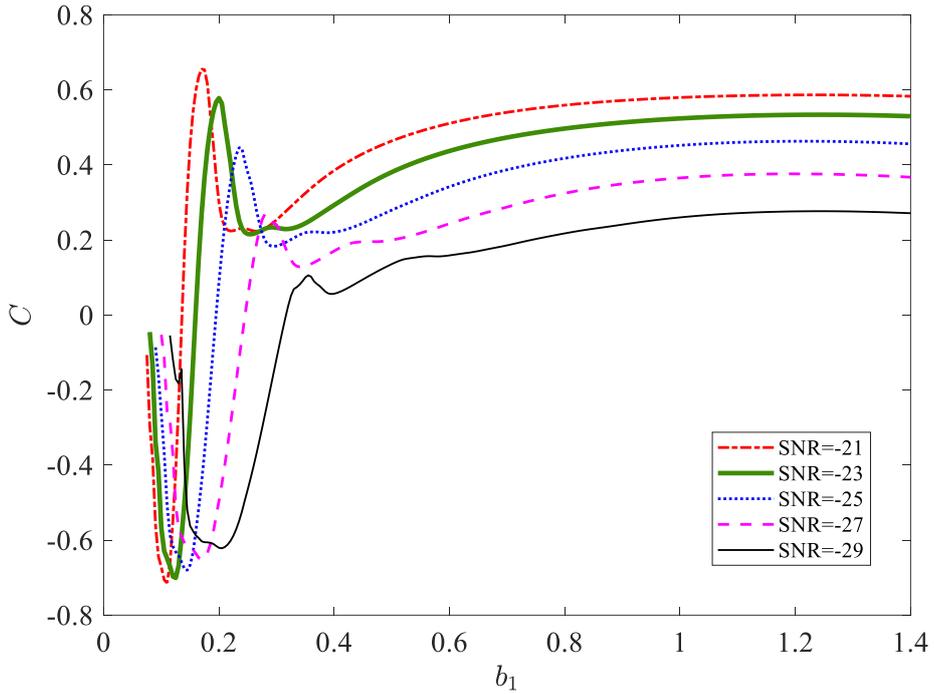

**Fig. 13.** The curve of the cross-correlation coefficient $C$ varies with the parameter $b_1$ when inputting different noise intensity.

### 4. Discussion

In the SR theoretical framework, noise is utilized noise to amplify the weak characteristic signal. It offers an advantage in extracting subtle features under strong noise background. Other processing techniques, while may also be effective in removing noise, but often come at the expense of compromising valuable weak feature information. In section 1, we introduce the classic non-stationary signal processing algorithms, including orthogonal wavelet transform, moving average filtering, and variational mode decomposition (VMD). By contrasting these methods with the SR approach under various noise conditions, we demonstrate that the SR method proposed in this paper yields a superior SNR.

Table 1 Comparison of nonstationary signal processing methods (The unit in the table is dB.)

| Sample | Input SNR | Wavelet | Moving average | VMD | SR |
| --- | --- | --- | --- | --- | --- |
| 1 | -16 | -17.1140 | -20.5786 | -17.2949 | -14.6545 |
| 2 | -20 | -20.3399 | -23.8031 | -20.6533 | -18.6514 |
| 3 | -24 | -24.5547 | -28.0268 | -24.7500 | -22.9044 |
| 4 | -27 | -28.4750 | -35.4960 | -28.5426 | -25.2907 |

Time varying scale is a method that the scale coefficient changes with time according to the characteristics of signal frequency. It has good adaptability to non-stationary signals. Some scholars have achieved good results in studying LMF in relatively low-frequency range, segmented LMF and other types of non-stationary signals by using stochastic resonance method [51, 52].

When dealing with other types of signals, the main difficulty is to choose an appropriate nonlinear model. The commonly used SR models include potential function with monostable, bistable, tristable and multistable configurations. However, different models have different adaptability to different signals. For example, monostable models can better adapt to impulse signals. When dealing with uncommon or mixed signals, researchers need to compare different models to carry out targeted research on their adaptability.

This method enhances the millimeter wave radar signal and promotes the development of radar signal processing. If algorithm is implanted into the radar mixer hardware circuit, which will effectively improve the accuracy of radar detection and reduce the subsequent processing time.

## 5 Conclusion

The traditional re-scaled method cannot enhance the UHF-LFM signal by SR. This paper solves the mentioned problem. A time-varying re-scaled method is established to study the MSR and ISR in an asymmetric bistable system subjected to the UHF-LFM signal excitation. The main conclusions are as follows:

(1) MSR and ISR of an asymmetric bistable system under the excitation of UHF-LFM signal are analyzed from the perspective of cross-correlated coefficient and phase trajectories. At the same time, the resonance regions include the bistable region, the monostable region and the ISR region. Among them, some resonance regions separate in the right tilted system and collapse in the left tilted system.

(2) Comparing the output of different resonance regions, the strongest resonant output in nonlinear system occurs in the bistable resonance regions. A set of SR region and ISR region excited by the same potential well have consistent outputs. Besides, the outputs of SR and ISR excited by different potential wells have the same time-frequency domain output and opposite time-domain output.

(3) The optimization of time-varying re-scaled cardinality can adjust the SR within a small parameter range, which mainly involves bistable resonance. Furthermore, the left tilted system reduces the generation of monostable region and ISR, ultimately improving the continuity of the signal time-frequency curve and enhancing the signals.

The proposed method realizes the strongest SR under the ultra-high frequency excitation. It also explains the mechanism and law of MSR and ISR of the asymmetric bistable system under the UHF-LFM signal excitation. The results are valuable for the study of complex response of a nonlinear system under the excitation of UHF-LFM signal.

## Acknowledgements

The project was supported by the National Natural Science Foundation of China (Grant No. 12072362), the Priority Academic Program Development of Jiangsu Higher Education Institutions, the Spanish State Research Agency (AEI) and the European Regional Development Fund (ERDF, EU) under Project No. PID2019-105554GB-I00 (MCIN/AEI/10.13039/501100011033).

## Data Availability Statement

There is not associated data proposed for this manuscript.

## Declarations

**Conflict of interest** The authors declare that they have no known competing financial interests or personal relationships that could have appeared to influence the work reported in this paper.